\documentclass[conference]{IEEEtran}
\IEEEoverridecommandlockouts
\usepackage{cite}
\usepackage{amsmath,amssymb,amsfonts}
\usepackage{algorithmic}
\usepackage{graphicx}
\usepackage{textcomp}
\usepackage{xcolor}
\usepackage{hyperref}
\usepackage{lipsum}
\usepackage{stfloats}
\usepackage{tabularx}

\usepackage[margin=0.75in]{geometry}
\usepackage{xspace}
\newcount\Comments  % 0 suppresses notes to selves in text
\usepackage{color}
\definecolor{darkgreen}{rgb}{0,0.5,0}
\definecolor{purple}{rgb}{1,0,1}
\definecolor{teal}{rgb}{0,0.4627,0.5804}
\newcommand{\kibitz}[2]{\ifnum\Comments=1\textcolor{#1}{#2}\fi}

\def\BibTeX{{\rm B\kern-.05em{\sc i\kern-.025em b}\kern-.08em
T\kern-.1667em\lower.7ex\hbox{E}\kern-.125emX}}

\begin{document}
\pagenumbering{arabic} 
\pagestyle{plain}

\newgeometry{left=1.91cm,right=1.91cm,top=2.54cm,bottom=1.91cm}

\title{Using Automated Vehicle Data as a\\Fitness Tracker for Sustainability}
% \title{Understand your Autonomic Vehicle as easy as using Apple
% Fitness}

\author{\IEEEauthorblockN{Xia Wang}
\IEEEauthorblockA{\textit{Department of Computer Science} \\
\textit{Vanderbilt University}\\
Nashville TN 37235, USA \\
xia.wang@vanderbilt.edu}
\and
\IEEEauthorblockN{Sobenna Onwumelu}
\IEEEauthorblockA{\textit{Department of Computer Science} \\
\textit{Vanderbilt University}\\
Nashville TN 37235, USA \\
sobenna.p.onwumelu@vanderbilt.edu}
\and
\IEEEauthorblockN{Jonathan Sprinkle}
\IEEEauthorblockA{\textit{Department of Computer Science} \\
\textit{Vanderbilt University}\\
Nashville TN 37235, USA \\
jonathan.sprinkle@vanderbilt.edu}
}

\maketitle

\begin{abstract}
This work describes the use of on-board vehicle data from cars with advanced driver assistance features as a trip summary, with the goal of helping drivers contextualize their driving habits in terms of sustainability. 
The approach is similar to recent advancements in fitness tracking apps, which leverage smartwatches and other wearable devices to characterize activities during a workout or as part of daily fitness monitoring. 
Instead of adding new vehicle sensors, the data used for this work is from on-board driving data, namely, signals decoded from the vehicle's Controller Area Network (CAN) bus.
With the deepening research of automatic driving technologies, Autonomous Vehicles (AVs) have gradually entered the consumer field, and more users are benefiting from the convenience and safety assistance provided by driving assistance and autonomous driving. 
However, various technical obstacles persist due to the complex environment, the non-communication of technologies, and users' trust. 
We propose indicators for evaluating the key characteristics of each drive, to facilitate drivers' familiarity with advanced driver assistance systems and to allow them to consider how different driving styles affect sustainability metrics. 
Further extensions will allow users to add feedback as part of the driving summary, laying a data foundation for future controller iterations based on real driving data and the attitude of drivers towards it.
\end{abstract}

\begin{IEEEkeywords}
Controller Area Network, Sustainability, Advanced Driver Assistance Systems, Energy, Safety
\end{IEEEkeywords}

\section{Introduction}

The research on personalized data tracking applications, such as those related to fitness monitoring, health assistance, and weight loss self-tracking, indicates that users can benefit from understanding key indicators of tracking and interacting with the tracked data. 
This interaction helps users grasp the current status of relevant indicators, which could further influence their behaviors and ultimately aid them in achieving their goals. 
Interactive behaviors with tracking apps include real-time awareness of the current status of monitored indicators, reviewing historical trends of indicators~\cite{jin2022self}, and receiving positive feedback on setting and consistently achieving reasonable goals related to key indicators. Simultaneously, the social factors of using tracking apps can also influence users' attitudes towards the monitored objects. By sharing their tracking achievements, users can further achieve a ``group progress'' effect~\cite{niess2021don}.

In this work, we introduce the inclusion of analogous self-tracking dashboard features in vehicle automation. 
If drivers are presented with key metrics such as safety, fuel efficiency, and comfort after each drive and compare the driving performance between manual control and the Adaptive Cruise Control (ACC),
%autonomous cruise controller (ACC) control,
the
comparison may allow them to assess the reliability and satisfaction with the equipped ACC. 
Additionally, drivers can access historical data, trends, and detailed charts for better understanding, adjusting ACC usage to suit their driving habits and varying road conditions. 
Furthermore, drivers can set sustainable goals, such as achieving a safety score over 95\%, fostering a positive loop to habitually use autonomous driving assistance. 
They can also compare current metrics with those of previous drives and historical averages to perceive corresponding rewards for using autonomous driving assistance. 
Finally, this technology can extend the social aspect by showcasing messages like ``your fuel efficiency index surpasses 89\% of drivers,'' expanding the application's influence. 
Such features are poised to enhance drivers' comprehension of the self-driving experience and to foster increased trust in cruise control among human operators.

In this work,  the proposed methodology offers the following contributions that we claim:
\begin{itemize}
    \item A sustainability dashboard tailored for connected and automated vehicles that furnishes key metrics encompassing safety, fuel efficiency, and passenger comfort, the trends of metrics and the comparison between the current values against historical records of such metrics.
    \item This approach utilizes low-latency, low-probability-of-anomaly signals from the CAN bus equipped on various vehicle models to offer a real-time, straightforward, and universal method for extracting key driving indicators.
    \item This work establishes a data format for the interaction between drivers and key indicators of AVs, providing a data foundation for the sustainable advancement of driver acceptance of driving assistance technologies, and for the iterative development of safer, more economical, and more comfortable autonomous driving technologies.
\end{itemize}

\section{Related Work}

\subsection{Sustainability in Self-driving Development}

The sustainability of the autonomous driving ecosystem is
\newgeometry{left=1.91cm,right=1.91cm,top=1.91cm,bottom=1.91cm}
\noindent reflected not only in the continuous iteration of technology but, more importantly, in considering human involvement. 
Encouraging users of self-driving, including drivers, passengers, and the general public, to enhance their understanding and trust in autonomous driving technologies through proper promotion, education, assistive technologies, etc., is an equally important issue. 
This helps people benefit more effortlessly and frequently from autonomous driving technologies. D.D. Heikoop et al.~\cite{heikoop2020practitioner} 
emphasized the importance of considering motivational elements as key human factors in the development of Advanced Driver Assistance Systems (ADAS).
It is also indispensable to consider usability and the enjoyment of driving for the implementation of human-machine interfaces (HMIs) of ADAS. 
Meanwhile, certain research concentrates on the macro-level sustainability implications of traffic control, last mile supply management, resource allocation, clean energy and effects to climate~\cite{XU2023203,10414920,sun2023resource,cheng2023estimating,fan2023combined}, the requirements of the Intelligent Transportation System (ITS) and vehicle technologies for the sustainable iteration of road transport infrastructure~\cite{rampalli2020redesigning}, and research on the contribution of Mobility-as-a-Service (MaaS) to sustainability~\cite{de2020mobility}.

\subsection{Visualization for Vehicle Performance}
Features presented in dashboard form could provide an interactive approach to help drivers better understand and learn faster about the functions of cruise control equipped cars, thereby increasing the transparency of the actions of cruise controllers. 
For instance, Yi~\cite{yi2022exploring} introduced a dashboard design used during the operation of autonomous vehicles, aiming to assist drivers in understanding the decision intentions of autonomous vehicles and enhancing the safety performance of these vehicles. 
The best interface design for AVs, a challenge in 2016~\cite{benderius2017best}, 
proposed that the design aspects of dashboard icons, including their shape, size, placement, and color, significantly influence the effectiveness of the interactive interface's display.
The most related work is~\cite{fugiglando2017characterizing}, which utilized CAN bus messages as a data source to analyze the ``drivers' DNA'' from 
elements of driving such as cautious braking, vigilant turning, maintaining safe speeds, and fuel-related energy efficiency.
Yet the focus of the ``drivers' DNA'' work is to horizontally compare characteristics among different drivers based on key driving metrics. 
In contrast, our paper places greater emphasis on drivers reviewing their own core driving statistics shortly after each drive, to make individual decisions such as determining which driving scenarios are more suitable for self-driving and whether to increase the proportion of autonomous driving control.

\subsection{Evaluations of AV Controllers}
In this work, we focus on three key metrics of performance evaluation of AVs, including safety, fuel efficiency and comfort.

\subsubsection{Safety}
Safety assessments of autonomous driving vehicles in the early 90s primarily focused on constructing collision models. These models aimed to estimate the probability of collisions involving autonomous vehicles with leading vehicles in traffic, as well as the likelihood of collisions with multiple vehicles. For instance, as illustrated in~\cite{touran1999collision,an2023runtime}, 
various collision prediction models were presented to estimate the likelihood of rear-end accidents under a range of conditions.
These models took into account factors such as the spacing between two vehicles, their pre-collision velocity, the variance in deceleration rates, and the latency in response times.
However, these probability models involve numerous assumptions and complex calculations, leading to potential errors. 
In this paper, we choose safety metrics that are widely embraced, simpler to compute, and rely on fewer assumptions, thus following metrics are under consideration: 

\begin{itemize}
    \item \textbf{Time Headway}: Time headway, or simply headway, was first presented by %Dr. 
    Drew in 1968~\cite{drew1968traffic}. Since then, many researchers continue to use this indicator to assess the safety aspect of driving vehicles~\cite{fenton1979headway} and provide comprehensive analyses~\cite{maurya2015speed} of headway in various driving scenarios. 
    Headway refers to the temporal distance between the leading car and the ego car, evaluating the secure interval between the two vehicles.
    Headway is crucial for maintaining safe following distance and avoiding rear-end collisions. 
    \item \textbf{Time to Collision (TTC)}: TTC was first introduced by Hayward in 1972~\cite{hayward1972near} and extensively discussed 15 years later~\cite{hyden1987development}. 
    TTC measures the projected duration before a vehicle collides with an object or another vehicle, assuming the current velocity and path are unchanged.
    It essentially gauges how close the vehicle is to a potential collision. A shorter TTC indicates a more imminent danger of collision, making it particularly valuable for assessing immediate collision risk.
\end{itemize}

\subsubsection{Fuel Efficiency}
Based on the signals provided by CAN bus gathered with Strym, outright calculating fuel efficiency is impossible, as fuel consumption was not available on the CAN network. Many formulas, such as one popularized by Wong~\cite{wong2001theory} that models fuel consumption, could be further converted to the estimation of fuel efficiency, with the following assumptions:
\begin{itemize}
    \item Power is roughly proportional to fuel consumption, as modeled by Song et al.~\cite{song2009estimation} in their exploration of fuel consumption and Vehicle Specific Power (VSP).
    \item Parameters such as vehicle mass and drag coefficient remain constant.
    \item Transmission or gear is not considered in this context.
    \item Increased consumption from starts and stops in traffic depends on driver behaviors.
\end{itemize}

Future work may take advantage of semi-principled fuel models that have recently been published, such as in \cite{khoudari2023reducing}.

\subsubsection{Comfort}
One of the primary concerns regarding autonomous vehicles is ensuring a comfortable ride, as a less comfortable experience could lead to decreased acceptance of this technology among the general public. In previous work~\cite{bellem2016objective,yan2021comfort}, 
factors like the driver's preferred style of driving, the conditions of the driving environment, and the speed of the vehicle play a role in determining the comfort level of vehicle drive.
For instance, passengers tend to be more sensitive to acceleration and changes in acceleration when traveling at high speeds on highways compared to driving on rural roads. While studying these factors could offer more meticulous comfort evaluation criteria, 
the main goal of this article is to create a more widely applicable metric.
Thus, we plan to place greater emphasis on vehicle motion factors that have a direct impact on the comfort metric of the driving experience~\cite{irmak2021individual,de2022relating}. In this context, vehicle motion can be categorized into two main types: lateral motion and longitudinal motion. Since this paper primarily focuses on an autonomous speed controller, specifically its operation in car-following mode, our attention will be directed towards assessing the impact of longitudinal vehicle motion on the comfort metric. 
These movements are usually short-lived and can be depicted as distinct bursts, which can be defined by their acceleration and jerk, linked by the frequency of the motion~\cite{de2020role,de2023standards}.

To evaluate the comfort of longitudinal vehicle motion,
we can turn to official standards and specifications. 
For instance, ISO 22179~\cite{isointelligent} outlines acceleration limits for Full Speed Range Adaptive Cruise Control systems, ranging from +4/-5 $m/s^2$ to +2/-3.5 $m/s^2$, along with negative jerk limits of -5/-2.5 $m/s^3$ under a speed of 20 m/s and 5 m/s, respectively. 

\subsection{Acquiring Data from AVs}
We use data from field experiments~\cite{bunting2022data}, which are publicly shared on CyVerse~\cite{swetnam2018cyverse}, 
a center dedicated to data science for storing and processing data. Larger scale data collection experiment can be found in ~\cite{lee2024traffic}.
The experimental dataset includes both CAN bus message data and GPS data gathered with LibPanda \cite{bunting2021libpanda}, and the data employed for creating the proposed dashboard primarily comes from parsed CAN bus data. Through the use of on-board data, high-frequency vehicle information, from which energy use, following distance, and other metrics can be extracted with sufficient accuracy to draw research conclusions.

\section{User-friendly Dashboard Design}

\subsection{Overview}
The  dashboard is structured into four distinct sections, as shown in Fig.~\ref{fig:dashboard}, each serving a specific purpose:

\begin{figure}[t]
    \centering
    \includegraphics[width=.45\textwidth]{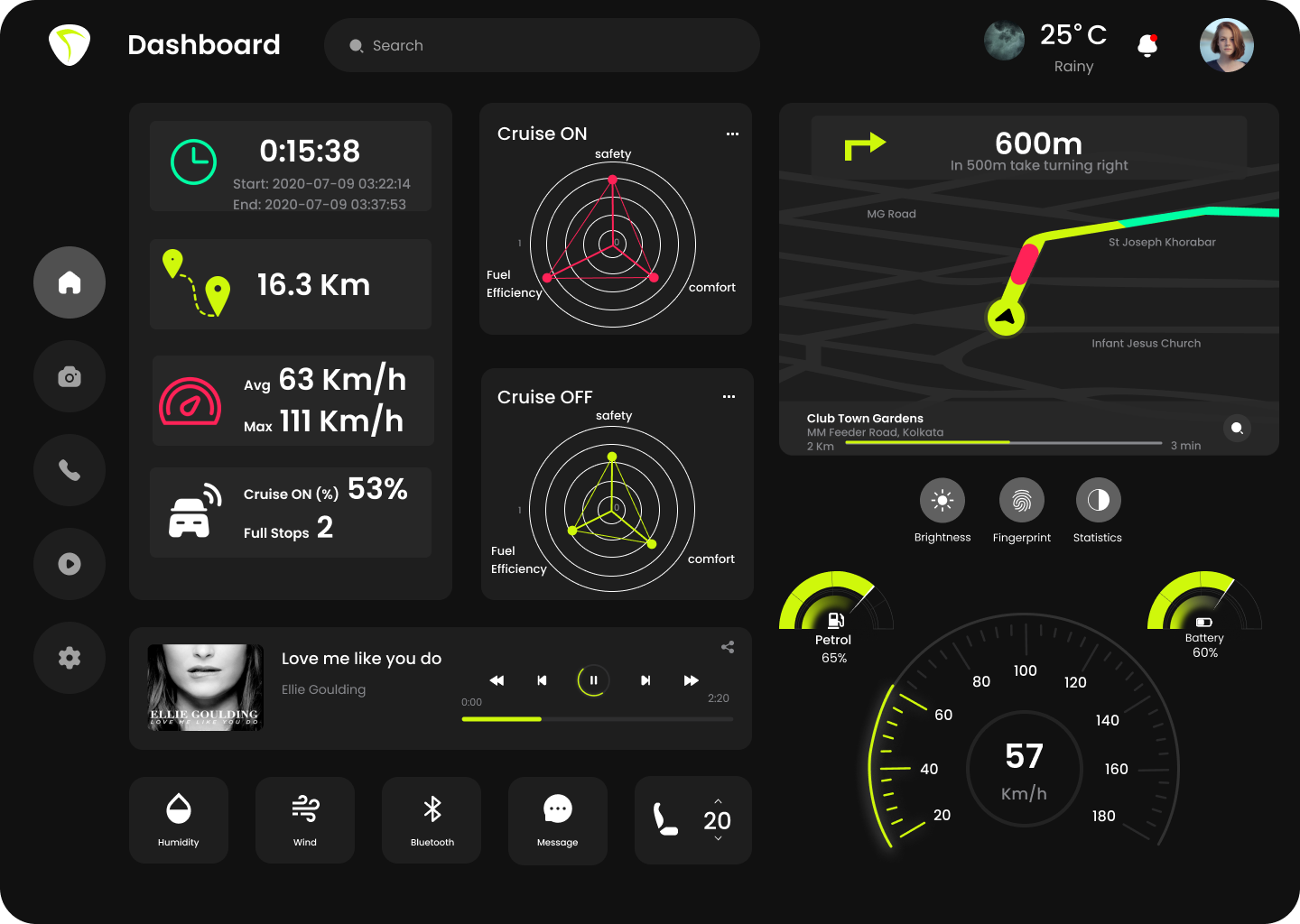}
    \caption{Overview of the proposed dashboard.}
    \label{fig:dashboard}
\end{figure}

\begin{itemize}
    \item In the upper-left corner, you will find the autonomous vehicle Key Performance Indicator (KPI) display area.
    \item Positioned in the upper-right corner is the navigation section, providing essential route information.
    \item On the lower-right side, we have the instrument panel area, presenting critical vehicle data.
    \item Finally, the functional area is fully customizable to cater to the driver's preferences. This section allows users to display current background music and entertainment features, as well as control and display common auxiliary functions such as air conditioning, seat adjustments, and Bluetooth connectivity.
\end{itemize}

UI design file is available through Figma\footnote{\url{https://www.figma.com/file/CRXoY137VGyROMDt7DATbh/AVFit?type=design&t=5ypMxtG9FXwE1DpD-6}}, and the code for calculating various metrics is available on GitHub\footnote{\url{https://github.com/Summer72Wang/Sustainability_Fitness_Tracker_for_AVs.git}}. 
The remainder of this section focuses on the design of the KPI area, which can be further divided into two key components:
\begin{itemize}
    \item Displaying fundamental aggregated indicators to answer "how long," "how far," and "how fast".
    \item Employing spider charts to graphically represent the effects of ACC on safety, fuel efficiency, and comfort.
\end{itemize}

\subsection{Key Metrics}
\subsubsection{Safety}
As mentioned above, both headway and TTC can provide safety estimation for self-driving rides. However, based on previous works~\cite{vogel2003comparison,ramezani2020comparing}, which compare headway and TTC in safety evaluation, we use headway as the safety metric in this paper for the following reasons:
\begin{itemize}
    \item Headway and TTC exhibit a meaningful positive correlation. A small headway often coincides with a reduced TTC, serving as an indicator of danger. Nevertheless, in braking situations, when the ego car's speed falls below that of the leading car, the TTC value becomes infinite, rendering it irrelevant in that specific context. Conversely, headway continues to offer a meaningful assessment of potential hazards.
    \item Headway, indicating the potential risk of tailgating, was deemed particularly suitable for the context of this research. On the other hand, the shorter TTC represents an immediate danger and is better suited for real-time warning systems while driving, effectively alerting drivers to the existing hazards.
    \item Headway is easy to measure, which is a continuous function, while TTC is a piecewise function.
\end{itemize}

We calculate headway using~\eqref{equation_headway}, where $D_{leading}$ is the longitudinal spacing between the ego and the leading vehicle, and $V_{ego}$ is the current ego car's velocity. 
For comparison purpose, the calculation equation of TTC is provided by~\eqref{equation_ttc}, where the $V_{leading}$ is the current leading car's velocity, while the $D_{leading}$ and the $V_{ego}$ remain same as above. 

\begin{equation}
Headway = \frac{D_{leading}}{V_{ego}}
\label{equation_headway}
\end{equation}

\begin{equation}
TTC = 
\begin{cases}
\frac{D_{leading}}{V_{ego} - V_{leading}},  & \text{if $V_{ego} > V_{leading}$} \\
\infty, & \text{if $otherwise$}
\end{cases}
\label{equation_ttc}
\end{equation}

\begin{figure}[h]
    \centering
    \includegraphics[width=.45\textwidth]{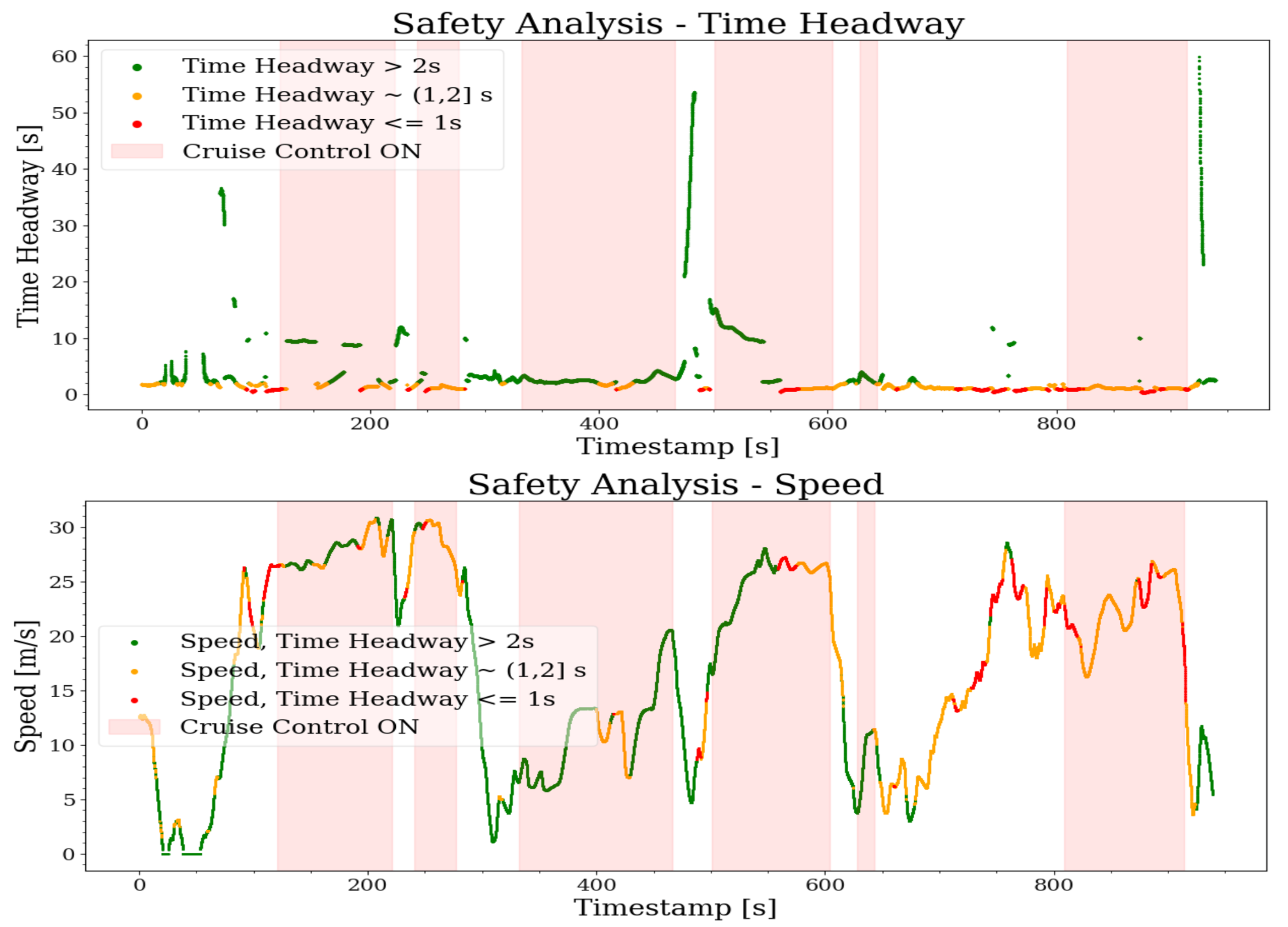}
    \caption{Safety analysis diagram showing the time headway values and speed values of categories of alert, attention and safe zones under conditions of cruise control ON and OFF.}
    \label{fig:safety_new}
\end{figure}

Finally, due to US follows 2 second rule of headway to recommend a minimum safe headway~\cite{hassan2016drivers}, we segment the headway time-series data into three categories: less than or equal to 1: the alert zone; greater than 1 and less than or equal to 2: the attention zone; and greater than 2: the safe zone. We subsequently compute the corresponding percentages for these three categories. The ultimate safety index value is determined by the proportion of the sum proportion of the safe zone and the attention zone. The more comprehensive visual information is shown in Fig.~\eqref{fig:safety_new}.

\subsubsection{Fuel Efficiency}

With \textit{$\Delta D$} being the change in driving odometer, the general formula represents the inverse relationship between fuel efficiency (\textit{FE}) and fuel consumption (\textit{FC}) as below~\eqref{eq:fe}: 

\begin{equation}
FE = \frac{\Delta D}{FC}
\label{eq:fe}
\end{equation}

Then, with \textit{FCR} being the rate of fuel consumption, \textit{FC} is calculated by~\eqref{eq:fc}. Also, notice that dividing by 100 here is because people usually measure fuel consumption in units of \textit{L/100km}.

\begin{equation}
FC = \frac{\Delta D * FCR}{100}
\label{eq:fc}
\end{equation}

The algorithm of calculating FCR is provided by~\eqref{eq:fcr}, where $V_{ego}$ is the ego car's velocity and the $ACC_{ego}$ is the ego car's acceleration. Also, the meaning of the parameters of \textit{a}, \textit{b}, \textit{c}, \textit{d} is provided below:

\begin{equation}
FCR = a + b \times {V_{ego}} + c \times {V_{ego}}^2 + d \times {ACC_{ego}}
\label{eq:fcr}
\end{equation}

\begin{itemize}
    \item \( \textbf{a} \)  is the base fuel consumption when the car is idling, given as 5, indicating the base fuel consumption is 5 \textit{L/100km} when idling.
    \item \( \textbf{b} \)  represents the linear term of rolling resistance, given as 0.05, indicating the rolling resistance coefficient is 0.05 \textit{L/100km} per \textit{kph}.
    \item \( \textbf{c} \) represents the quadratic term of aerodynamic drag, given as 0.001, indicating the aerodynamic drag coefficient is 0.001 \textit{L/100km} per ${(kph)}^2$.
    \item \( \textbf{d} \)  represents the effect of acceleration, given as 0.2, indicating the effect of acceleration coefficient is 0.2 \textit{L/100km} per ${(m/s)}^2$.
\end{itemize}

\begin{figure}[h]
    \centering
    \includegraphics[width=.45\textwidth]{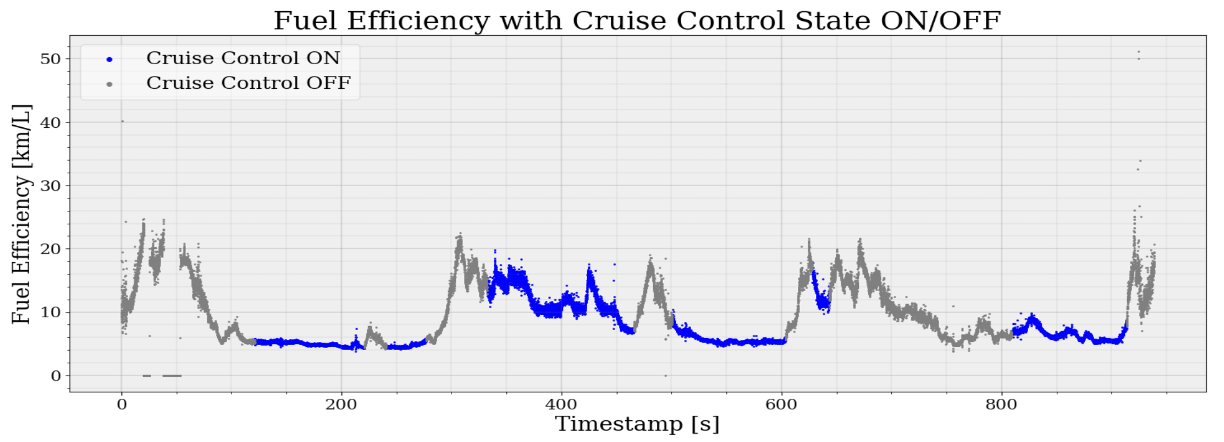}
    \caption{Fuel efficiency diagram showing the efficiency plotted as cruise control state changes.}
    \label{fig:fuel_effi}
\end{figure}

In short, by considering the predicted \textit{FCR} with changes in distance given by odometer readings, the estimation accounts for fuel efficiency inside and outside of cruise control, shown in Fig.~\ref{fig:fuel_effi}. We use a \textit{MinMaxScaler} to convert the real fuel efficiency values into the fuel efficiency index for each ride.

\subsubsection{Comfort} 
We use acceleration (A) and jerk (J) to evaluate the comfort metric of a cruise control vehicle. 
Jerk is the rate of change of acceleration over time defined in~\eqref{equation3}.
\begin{equation}
J = \frac{\Delta A}{\Delta t}
\label{equation3}
\end{equation}
 We evaluate the jerk data to determine how smoothly the vehicle's acceleration is changing. A lower jerk value indicates smoother and more comfortable acceleration transitions, while a higher jerk value suggests more abrupt changes in acceleration, potentially leading to a less comfortable ride. 
The majority of jerk values are concentrated within 5 $m/s^3$ with some large outliers.

\begin{table*}[ht]
\centering
% \tiny
% \scriptsize
\footnotesize
\caption{Comparisons on key metrics. ON and OFF describe the Cruise Control state.}
\begin{tabular}
% {\textwidth}
{|l|lll|lll|lll|lll|l|}
\hline
\multicolumn{1}{|c|}{%\multirow{2}{}{}
    {}} &
  \multicolumn{3}{c|}{\textbf{Safety Index (\%)}} &
  \multicolumn{3}{c|}{\textbf{Fuel Effic. Index (\%)}} &
  \multicolumn{3}{c|}{\textbf{Fuel Effic. (km/L;\%)}} &
  \multicolumn{3}{c|}{\textbf{Comfort Index (\%)}} &
  \multicolumn{1}{c|}{%\multirow{2}{*}
  {\textbf{ACC ON \%}}} \\ \cline{2-13}
\multicolumn{1}{|c|}{} &
  \multicolumn{1}{c}{\textbf{ON}} &
  \multicolumn{1}{c}{\textbf{OFF}} &
  \multicolumn{1}{c|}{\textbf{All}} &
  \multicolumn{1}{c}{\textbf{ON}} &
  \multicolumn{1}{c}{\textbf{OFF}} &
  \multicolumn{1}{c|}{\textbf{All}} &
  \multicolumn{1}{c}{\textbf{ON}} &
  \multicolumn{1}{c}{\textbf{OFF}} &
  \multicolumn{1}{c|}{\textbf{All}} &
  \multicolumn{1}{c}{\textbf{ON}} &
  \multicolumn{1}{c}{\textbf{OFF}} &
  \multicolumn{1}{c|}{\textbf{All}} &
  \multicolumn{1}{c|}{} \\ \hline
\textbf{Avg. of nearest 5 rides} &
  100.0 &
  98.8 &
  98.8 &
  31.9 &
  14.9 &
  22.7 &
  8.19 &
  6.26 &
  7.03 &
  95.3 &
  91.8 &
  92.1 &
  37.3 \\
\textbf{Previous ride} &
  100.0 &
  94.0 &
  94.2 &
  17.6 &
  34.1 &
  26.1 &
  6.76 &
  8.41 &
  7.61 &
  100.0 &
  90.8 &
  91.1 &
  3.1 \\
\textbf{Recent ride} &
  90.2 &
  84.7 &
  87.6 &
  24.9 &
  61.2 &
  40.1 &
  7.49 &
  11.12 &
  9.01 &
  90.2 &
  87.9 &
  89.1 &
  52.6 \\
\textbf{Change rate (to avg.)} &
  -9.8 &
  -14.3 &
  -11.3 &
  -21.9 &
  311.8 &
  76.5 &
  -8.55 &
  77.75 &
  28.13 &
  -5.3 &
  -4.3 &
  -3.3 &
  41.0 \\
\textbf{Change rate (to prev.)} &
  -9.8 &
  -9.9 &
  -7.0 &
  41.5 &
  79.5 &
  53.6 &
  10.80 &
  32.22 &
  18.40 &
  -9.8 &
  -3.2 &
  -2.2 &
  1596.8 \\ \hline
\end{tabular}%
\label{table:comp}
\end{table*}

\begin{figure}[h]
    \centering
    \includegraphics[width=.45\textwidth]{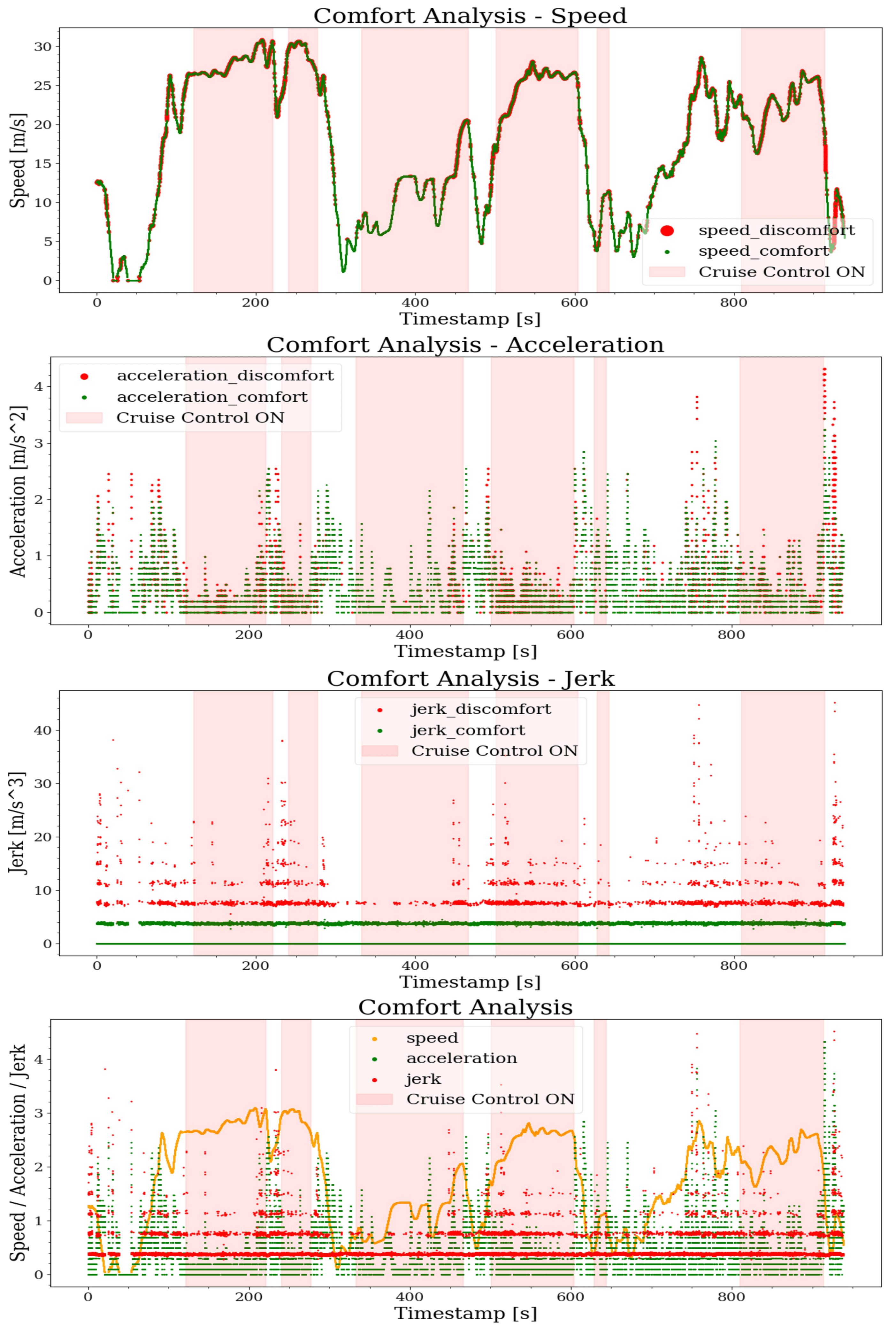}
    \caption{Comfort analysis diagram showing the speed, absolute acceleration, absolute jerk values of categories of comfort and discomfort under conditions of cruise control ON and OFF.}
    \label{fig:comfort_new}
\end{figure}

Further, we establish discomfort as a specification where the values of acceleration or jerk surpass specified thresholds. Undoubtedly, considering variations in occupant tolerance for discomfort during vehicle speeding up and slowing down, as well as the divergent levels of tolerance at high and low speeds, would enhance the robustness of the comfort metric. However, it is imperative to conduct comprehensive field experiments or simulation studies to meticulously measure these critical thresholds, an area warranting further investigation in subsequent research endeavors. In this paper, we utilize the \textbf{property}: $\square_{[1,n]} ( (A > 2) \vee (A < -3.5) \vee (J > 5) \vee (J < -5) \rightarrow \mathbf{discomfort} )$ to define discomfort and subsequently compute the proportion of comfortable driving. This proportion serves as a quantitative representation of the comfort index within the radar chart. 
The more intricate details are shown in Fig.~\ref{fig:comfort_new}.

\subsection{Trend and Comparison}
\begin{figure}[h]
    \centering
    \includegraphics[width=.4\textwidth]{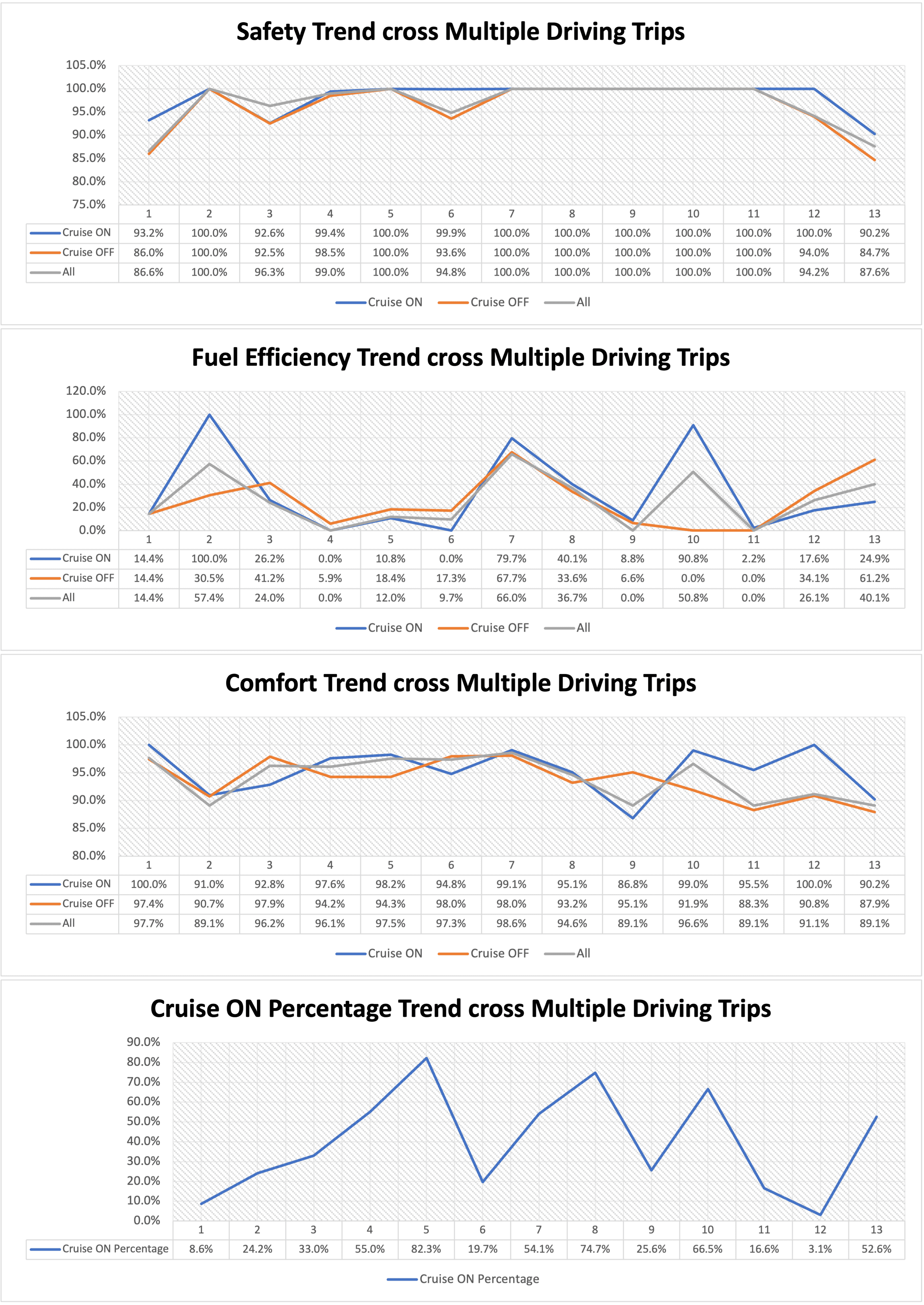}
    \caption{Trends of the safety index, fuel efficiency index, comfort index and cruise control ON percentage.}
    \label{fig:trend_all}
\end{figure}

Finally, we store driving records to display the time-series trends of various statistics, as well as a comparison of a current trip with the previous ones. 
Drivers are able to view side-by-side statistics of each ride on their own and to form their own analysis about their driving performance with conditions such as cruise control on or off. For instance, we show the trends of the three key metrics: safety, fuel efficiency and comfort, ae well as the cruise control on percentage trend in Fig.~\ref{fig:trend_all}. Also, we show the comparisons between the recent ride with the previous ones on these metrics in Table~\ref{table:comp}.

\section{Conclusion}

This project explored the usage of Strym's CAN bus data decoding possibilities and used them to create a coherent visualization system. 
Admittedly, the discussion in this paper has certain limitations. For instance, in calculating the time headway, the leading distance is a crucial variable; however, not all vehicle models provide this signal for analysis and utilization. Additionally, the parameters used in the computation of the fuel efficiency indicator do not consider the uncertainty of external environmental variation. Beyond these limitations, we also plan to advance a series of foreseeable future works. Such efforts include providing a more comprehensive analysis of the correlation between cruise control status and core self-driving indicators, as well as an edge case analysis of drivers' mistrust in ACC, leading to manual disengagement of cruise control. Furthermore, for the indicators of autonomous driving comfort, the contribution of lateral signals can also be incorporated.
To sum up, display of information that is usually under-the-hood can not only help drivers to understand their driving habits but also save researchers valuable steps in analyzing data in ways before impossible to do with a glance.

\section*{Acknowledgment}
This project received funding from the National Science Foundation, provided through Grant No. 2151500.

%\bibliographystyle{IEEEtran}
%\bibliography{IEEEabrv,references}

\end{document}